# Li$_2$MnCl$_4$ single crystal: new candidate for red-emitting neutron scintillator

Vaněček Vojtěch[a,b,c*], Král Robert[a], Křehlíková Kateřina[a], Kučerková Romana[a], Babin Vladimir[a], Petra Zemenová[a], Rohlíček Jan[a], Málková Zuzana[a], Jurkovičová Terézia[a], Nikl Martin[a]

Novel red-emitting scintillator Li$_2$MnCl$_4$ is proposed as a candidate for thermal neutron detection. It features high Li content, low density, low effective atomic number, and emission in red-NIR region. These characteristics make it an interesting candidate for long distance neutron detection in harsh enviroments e. g. decomisioning of nuclear powerplants. The absorption spectrum is thoroughly investigated in the scope of Tanabe-Sugano diagram. Luminescence mechanism in the undoped Li$_2$MnCl$_4$ is studied in depth. Doping by Eu$^{2+}$ and Ce$^{3+}$ is introduced as a trial to improve the scintillation efficiency. We show in the Eu$^{2+}$ and Ce$^{3+}$ doped Li$_2$MnCl$_4$ that luminescence mechanism involves energy transfer from the dopants to Mn$^{2+}$, and propose the local lattice distortion around the dopant and a possible charge compensation mechanisms.

## Introduction

Detection and spectroscopy of neutrons is crucial for applications ranging from large spallation sources [1] to nuclear non-proliferation [2]. The $^3$He shortage [3] limits future use of $^3$He proportional counters which are currently the most commonly used neutron detectors. The main source of $^3$He is radioactive decay for tritium. After World War 2 both US and CCCP kept large reserves of tritium for military uses. Therefore, an abundance of $^3$He was available as byproduct of nuclear warfare. However, after the end of the cold war both USA and USSR significantly lowered their tritium reserves and, as a result, production of $^3$He significantly dropped. Decreasing production of $^3$He together with increasing demand results in inevitable depletion of $^3$He reserves and steep increase in the price which is becoming unaffordable for non-US industry. Therefore, it is necessary to develop novel neutron detectors, that would be able to replace $^3$He proportional counters in some of the applications. One of the fields were not only replacement of $^3$He detectors, but also advances in neutron detection are necessary are neutron scattering experiments at large spallation sources [4] and decommissioning of nuclear powerplants [5].

This work reports on the optical, luminescence and scintillation characteristics of Li$_2$MnCl$_4$ single crystal, undoped, Ce and Eu doped, which is an interesting candidate for thermal neutron detection due to high content of Li and low density which assures effective n/γ discrimination in mixed radiation fluxes which is the most frequent case in applications. Red-infrared emission is favourable for the use of modern semiconductor photodetectors.

## Experimental

As starting materials powders of LiCl (99+ %, Merck) and MnCl$_2$ (97 %, Thermo Fisher Scinetific), EuS (99.99 %, Changsha Easchem), and Ce$_2$S$_3$ (99.99 %, Changsha Easchem) were used. Both LiCl and MnCl$_2$ were further purified by introduction of halogenating agents (HCl and COCl$_2$) into the melt and successive zone refining (> 20 passes of the melted zone) according to [6,7]. Material from the middle section of the zone refined ingot was selected for crystal growth for both LiCl and MnCl$_2$. In the case of LiCl the zone refined ingot was transparent and colourless. In the case of MnCl$_2$ the zone refined ingot was of a deep red colour and formed flakes upon mechanical stress. Such behaviour is typical for divalent halides with layered crystal structure e. g. MgCl$_2$, CdCl$_2$, CdBr$_2$, CdI$_2$, or PbI$_2$. Stoichiometric amounts of purified LiCl and MnCl$_2$ (and EuS or Ce$_2$S$_3$) were weighted into fused silica ampoules and sealed under vacuum using oxygen-hydrogen torch. The crystals were grown using single zone inductively heated vertical Bridgman furnace as described in [8]. During the entire growth the furnace was continuously evacuated using rotary vane pump to minimize heat transfer via convection and therefore prevent unwanted chimney effect. This results in a steeper temperature gradient and more stable temperature profile. The pressure was maintained at approx. 1 Pa during the entire growth. The growth rate was 0.6 mm/h, and the cooling time was 40 hours.

For the X-ray powder diffraction (XRPD) analysis, the crystal samples were powdered in alumina mortar and pestle and placed in the Ø 0.5 mm borosilicate-glass capillary in an atmosphere-controlled glovebox (concentration of $O_2$ and $H_2O$ < 1 ppm). The capillary was sealed with rubber to prevent degradation of the sample during measurement. Powder diffraction data were collected using the Debye-Scherrer transmission configuration on the powder diffractometer Empyrean of PANalytical (λCu,Kα = 1.54184 Å) that was equipped with a focusing mirror, capillary holder, and PIXcel3D detector.

Radioluminescence (RL) spectra, measured in the spectral range of 190-800 nm at room temperature, were obtained using a custom-made spectrofluorometer 5000M, Horiba Jobin Yvon. Tungsten-cathode X-ray tube Seifert was used as the excitation source (at 40 kV, 15 mA). The detection part of the set-up consisted of a single grating monochromator and photon-counting detector TBX-04, Horiba or an Ocean Optical CCD detector (based on the spectral range). Measured spectra were corrected for the spectral response of the setup. Photoluminescence decays were measured at 5000M using multichannel scaling and time-correlated single photon counting techniques under the excitation by microsecond flashlamp (μs-ms time scale) and nanosecond nanoLED (sub- μs time scale) excitation sources, respectively. The measured decays were approximated by sum of exponential terms convoluted with the instrumental response to excitation pulse.

The non-isothermal DSC was carried out using Setaram Themys 24 apparatus. The nominal charge of 65 mg of the powder $Li_2MnCl_4$ material was sealed in a quartz ampule depending under vacuum. Sealing the $Li_2MnCl_4$ powder under vacuum in a small quartz ampule enabled to simulate and monitor heat conditions similar to those present during the $Li_2MnCl_4$ crystal growth by Bridgman method. All DSC experiments were measured with the heating rate 10 K/min in the temperature range 25–650 °C and with an empty sealed quartz ampule in alumina crucible as a reference. The DSC apparatus was calibrated in the temperature range of 25–1300 °C using following standards (In, Sn, Zn, Al, Ag, and Au). The standard deviation of performed calibrations was in the range of ± 1 K. Processing of the obtained data was carried out by the Calisto Processing software.

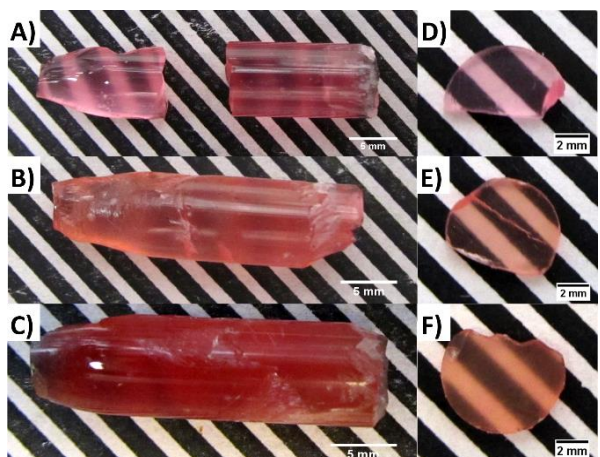

Figure 1: Photos of as grown crystals of A – C and samples prepared for optical characterization D – F of undoped $Li_2MnCl_4$ (A, D), $Li_2MnCl_4$:$Eu^{2+}$ (B, E), and $Li_2MnCl_4$:$Ce^{3+}$ (C, F) respectively.

The ratio of lithium and manganese in the $Li_2MnCl_4$ monocrystal samples was determined by flame atomic absorption spectrometry (FAAS). Samples with weight from 5.8 mg to 8.5 mg were slowly heated in 3ml HCl (1:1) (HCl purchased from P-Lab, Czech Republic) and fully decomposed. When the temperature of samples dropped, they were transferred into 100 ml volumetric flasks and 7 ml of KCl (with concentration 6 g of KCl in 100 ml) were added to decrease ionization of lithium. Finally, the distilled water was added to the volume of 100 ml in a volumetric flask. The calibration curve was prepared as combined for both elements using lithium and manganese standard solutions by the same way as samples solutions. A Varian AA240 FAAS (Varian Inc., Palo Alto, California, US) equipped with Li and Mn hollow cathodes (Varian Inc., Palo Alto, California, US) was used for the elements' concentration determination. The instrument conditions, such as wavelengths (670.8 nm for Li and 403.1 nm for Mn, respectively), 1.0 nm slit width, 10 cm burner, and the air-acetylene flame were employed according to the standard recommendations for both elements [9]. All the solutions were manually injected. The concentration of Li and Mn were determined from the concentration curves and subsequently the ratio of the Li and Mn in the samples was calculated. The relative standard deviation of determination was 2 %.

Due to the hygroscopic nature of used materials great care was taken to prevent degradation due to reaction with moisture. Whenever possible handling of all hygroscopic materials (starting materials, zone refined ingots, grown crystals etc.) was performed in the atmosphere-controlled glovebox using dry nitrogen (concentration of $O_2$ and $H_2O$ < 1 ppm) as the protective atmosphere. When manipulation outside of the glovebox was necessary the materials were placed inside fused silica ampoules closed with vacuum-tight valve or seal. However, due to technical limitation, most of the optical characterization was performed at ambient conditions. Whenever possible samples were immersed in/coated by luminescence free immersion oil to supress the degradation process.

**Results and discussion**

The crystal growth under optimized conditions resulted in homogeneous boules with no macroscopic defects. The undoped crystal is transparent and pink in colour while both $Eu^{2+}$ and $Ce^{3+}$ doped samples are deep red, which limits their transparency (see Fig. 1a-c). A cut and polished disc – like samples with dimensions Ø7x1.5 mm were prepared for all three crystals (see Fig. 1d-f).

The XRD measurement was performed on the powder $Li_2MnCl_4$ (LMC) samples (selected from the middle of the as-grown crystal) confirming the presence of $Li_2MnCl_4$ phase (see Fig. 2).

The $Li_2MnCl_4$ exhibits spinel crystal structure (space group Fd-3m, no. 227, phase prototype $MgAl_2O_4$). Using a general spinel formula:

$$(A_{1-x}B_x)^{tet}[A_{x/2}B_{1-x/2}]_2^{oct}X_4,$$

where elements in round brackets occupy tetrahedral lattice site, elements in square brackets occupy octahedral lattice site, x = 0 represents regular spinel, and x = 1 an inverse spinel structure a $Li_2MnCl_4$ can be represented as:

$$(Li)^{tet}[Li_{0.5}Mn_{0.5}]_2^{oct}Cl_4,$$

where A = $Mn^{2+}$, B = $Li^+$, X = $Cl^-$, and x = 1. Therefore, $Li_2MnCl_4$ is a fully inverse chloride spinel as reported by Lutz and Schneider [10]. Very weak reflections corresponding to LiCl were identified in the diffractograms of all three $Li_2MnCl_4$ crystals (see Fig. 2). It can be observed mainly through the two strongest reflections at 30.1 ° and 34.9 ° corresponding to the (1 1 1) and (2 0 0) lattice planes of cubic LiCl respectively. The presence of LiCl is probably due to a slight off-stoichiometry and will be a subject to a future crystal growth optimization.

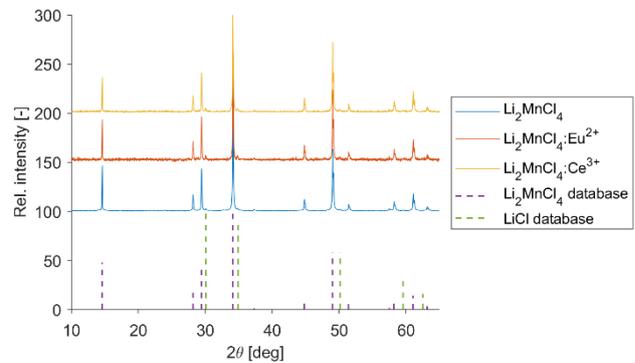

Figure 2: X-ray powder diffraction of samples prepared from all three $Li_2MnCl_4$ crystals.

To further elucidate on the composition of the grown crystals, the Li:Mn molar ratio was measured using flame atomic absorption spectroscopy (FAAS). Samples from the start (conic part) and the end (cylindrical part) of the grown undoped crystals were analysed. The FAAS results revealed that in the samples taken from the start (first-to-freeze) and the end (last-to-freeze) of the as grown undoped $Li_2MnCl_4$ the atomic ratio of Li:Mn is close to the ideal stoichiometry (see Tab. 1). However, the results point toward slight depletion of Li towards end of the growth. This could be caused by high lithium mobility in $Li_2MnCl_4$ crystals. The ionic mobility of $Li^+$ in several chloride spinels with composition $Li_2BX_4$ where B = Mn, Cd, Mg, Fe and X = Cl or Br [11–13] was investigated in 1980s for application as solid Li electrolytes. The temperature dependent conductivity measurements showed that the conductivity of the melt is only twice as high as conductivity of the solid sample just below the melting point [12]. Such a high ionic conductivity could result in significant $Li^+$ migration in the solid phase during crystal growth.

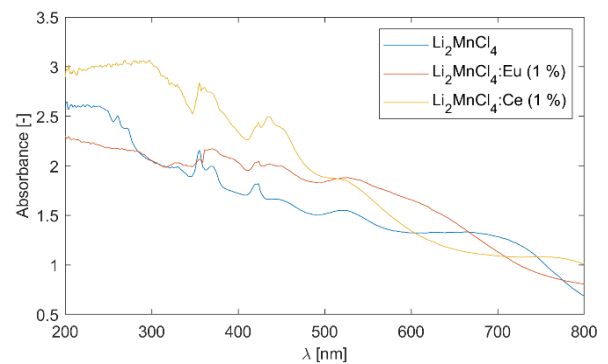

Figure 3: Absorption spectra of the $Li_2MnCl_4$ samples.

The DSC measurement revealed two endothermic effects with onset temperatures 442 ± 3 °C and 573 ± 1 °C (see Fig. S1a) on the heating curve and 459 ± 1 °C and 570 ± 1 °C (see Fig. S1b) on the cooling curve. Both effects were reproducibly detected in four heating-cooling cycles. Therefore, it is assumed that the effects are reversible. Endothermic effect around 450 °C was reported by several authors [11,14,15], Lutz et al.[15] ascribed this effect to the phase transition from spinel to the defect rock salt structure. Such phase transition might result in residual stress in the as grown crystals. However, no cracking of the crystals during processing was observed. The second effect at 573 ± 1 °C is ascribed to congruent melting of the high temperature $Li_2MnCl_4$ phase and corresponds well with the values reported in the literature [14,15]. Interestingly, undercooling of only 3 °C was observed. This effect is probably caused by the low volume to surface ratio of the sample and large contact surface of the melt with the quartz ampule which serves as heterogeneous nucleation centre.

Table 1: Results of the flame atomic absorption spectrometry. Samples labelled s were taken from

| Sample | $m$ [mg] | $m$(Mn) [mg] | $m$(Li) [mg] | Li:Mn |
|--------|----------|--------------|--------------|-------------|
| s1 | 8.1 | 2.12 | 0.54 | 2.02 : 1.00 |
| s2 | 6.5 | 1.71 | 0.42 | 1.95 : 1.00 |
| e1 | 8.5 | 2.2 | 0.54 | 1.95 : 1.00 |
| e2 | 6.8 | 1.78 | 0.43 | 1.91 : 1.00 |

the first-to-freeze section and samples labelled "e" were selected from the last-to-freeze section of the as grown undoped $Li_2MnCl_4$ crystal.

**Absorption spectra**

The measurement of absorption was complicated due to continuous degradation of the sample during measurement. The degradation results in formation of structured surface which results in strong scattering. Such scattering manifests as a diffuse background which is non-linearly increasing towards shorter wavelengths. Therefore, reliable estimation of the optical band gap from the absorption edge was impossible. This effect is also responsible for the atypical shape of the absorption spectra at longer wavelengths. Since the measurements proceeded from long wavelength side of the spectrum the beginning of the measurement is heavily affected by ongoing degradation of the sample. Nevertheless, distinct absorption bands could be observed for all three samples (see Fig. 3).

The absorption spectrum of the undoped sample was analysed in scope of Tanabe–Sugano diagram. The positions of the absorption bands were extracted via fitting. To suppress the influence of the diffuse background on the position of the absorption bands the first derivative of the absorption spectrum was fitted assuming gaussian shape of the absorption bands (see Fig. S2). The use of six gaussian components resulted in satisfactory fit in the range from 2 eV (620 nm) to 4 eV (310 nm). The vast majority of the experimental points are within 95 % prediction bounds of the used model. Although, the peak centred around 3.36 eV (370 nm) is quite asymmetric which results in lower accuracy of the fit, the goodness of fit is still satisfactory. The asymmetric shape of the absorption band might be caused by superposition of two bands. For illustration, Fig. S3 shows comparison between the measured absorption spectra and normalized spectrum reconstructed from the first derivative fitting. The positions of all six bands were compared to the d$^5$ Tanabe-Sugano diagram. The best agreement was at $D_q/B$ = 9.1 which corresponds to value of Racah parameter $B$ = 725 ± 7 cm$^{-1}$. The decrease of $B$ compared to free ion $B_{free}$(Mn$^{2+}$) = 860 cm$^{-1}$ [16] is due to nephelauxetic effect. The value $B/B_{free}$ of 0.84 is reasonable compared to 0.95 for MnF$_2$, 0.90 for NaCl:Mn$^{2+}$ [16], or 0.88 for CaCl$_2$:Mn$^{2+}$ [17].

Figure 4 depicts the absorption spectra of undoped Li$_2$MnCl$_4$ with indication of band positions and their assignment based on d$^5$ Tanabe-Sugano diagram. The double peak with maxima around 420 nm is due to coinciding $^6A_{1g}$ ($^6S$) → $^4A_{1g}$ ($^4G$) and $^6A_{1g}$ ($^6S$) → $^4E_g$ ($^4G$) transitions and no direct assignment between the two can be currently made. The mean maxima position was used for calculation of the $B$ Racah parameter.

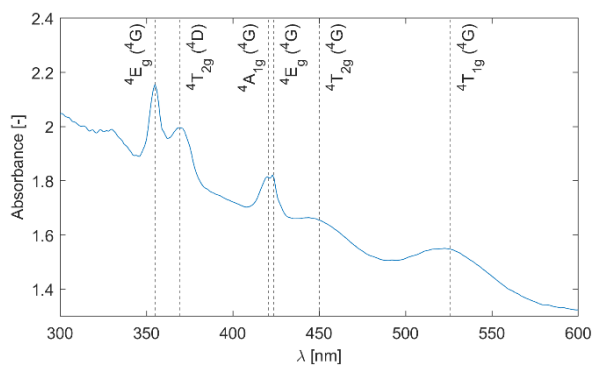

Figure 4: Absorption spectra of undoped Li2MnCl4 with indication of band positions and their assignment based on d$^5$ Tanabe-Sugano diagram.

Figure 5 depicts a section of the d$^5$ Tanabe-Sugano diagram with indication of the $D_q/B$ = 9.1. Transitions from the ground sextet to all quartets can be observed in the absorption spectra. The only exception is transition $^6A_{1g}$ ($^6S$) → $^4T_{1g}$ ($^4P$) which should be positioned around 320 nm. The position roughly corelates with a weak deformed peak at 330 nm. Therefore, we assume that the absorption band centred at 330 nm corresponds to the $^6A_{1g}$ ($^6S$) → $^4T_{1g}$ ($^4P$) transition. The PLE measurements are in agreement with this assumption (see Fig. 6 below). Transitions from the ground sextet to the doublet levels ($^2I$) exhibit very low oscillator strength due to high difference in spin number, i.e. strongly spin/forbidden transitions. Therefore, these transitions should not be observable. However, overlap of the $^6A_{1g}$ ($^6S$) → $^4T_{2g}$ ($^4D$) and $^6A_{1g}$ ($^6S$) → $^2T_{2g}$ ($^2I$) transitions might be the reason for the asymmetry of the $^6A_{1g}$ ($^6S$) → $^4T_{2g}$ absorption band.

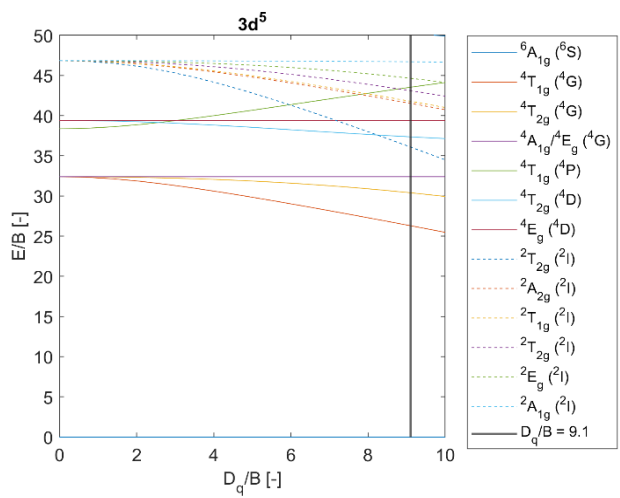

Figure 5: Section of the d$^5$ Tanabe-Sugano diagram with indication of the $D_q/B$ = 9.1. Doublet levels are in dashed lines. Reproduced from [18].

**Photoluminescence spectra**

Li$_2$MnCl$_4$

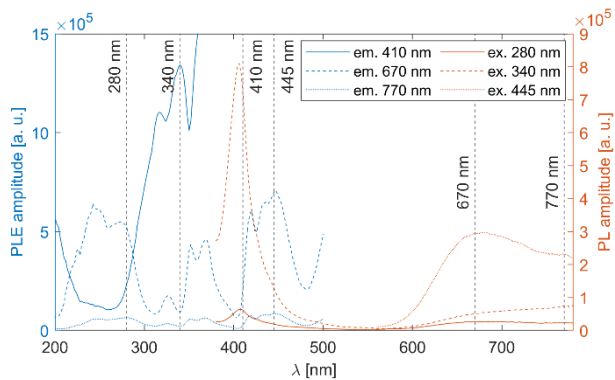

Figure 7: PLE and PL spectra of the Li$_2$MnCl$_4$:Eu$^{2+}$ (1 %) sample. The excitation and emission wavelengths are indicated by dashed lines. The high intensity around 200 nm is due to overcorrection.

The PL spectra (see Fig. 6) show broad band centred around 620 nm for all excitation wavelengths. This band can be ascribed to the $^4T_{1g}$ ($^4G$) → $^6A_{1g}$ ($^6S$) transition. Excitation to the higher quartet levels is followed by the non-radiative relaxation to the lowest excited level ($^4T_{1g}$) and radiative transition to the ground level ($^6A_{1g}$). Broad emission is due to different slopes of the $^4T_{1g}$ ($^4G$) and the $^6A_{1g}$ ($^6S$) levels at $D_q/B$ = 9.1. and sensitivity of the $^4T_{1g}$ ($^4G$) excited state to the crystal field. Increasing intensity towards 500 nm is due to the excitation light scattering. The PLE spectra monitored at 620 nm correspond well with the absorption spectra (see Fig. S4). The excitation band centred around 330 nm is much more pronounced compared to the absorption spectra which corroborates with assignment to the $^6A_{1g}$ ($^6S$) → $^4T_{1g}$ ($^4P$) transition.

### Li$_2$MnCl$_4$:Eu$^{2+}$

In PL of the Eu$^{2+}$ doped Li$_2$MnCl$_4$ sample (see Fig. 7) a narrow band cantered at 405 nm is observed. Based on the shape and position of the band[19] it is ascribed to 5d→4f emission of Eu$^{2+}$. Interestingly, the red emission band extends much further into NIR (outside of the sensitive region of used PMT) compared to undoped sample. Moreover, there is partial overlap between emission of Eu$^{2+}$ and excitation of the host matrix ($\lambda_{em}$ = 670 nm). Such overlap could enable energy transfer from Eu$^{2+}$ to Mn$^{2+}$ further facilitated by close proximity of both of them. Indeed, the energy transfer from Eu$^{2+}$ to Mn$^{2+}$ can be seen in the PL spectrum excited at 340 nm. Based on comparison of PLE spectra of undoped Li$_2$MnCl$_4$ (see Fig. 6, $\lambda_{em}$ = 620 nm) and Eu$^{2+}$ in Li$_2$MnCl$_4$:Eu (see Fig. 7, $\lambda_{em}$ = 410 nm), Eu$^{2+}$ is selectively excited at 340 nm. However, the PL spectrum of Li$_2$MnCl$_4$:Eu excited at 340 nm clearly shows both the Eu$^{2+}$ and Mn$^{2+}$ emissions (see Fig. 7). The energy transfer from Eu$^{2+}$ to Mn$^{2+}$ also gives rise to the additional excitation bands <320 nm in the PLE spectrum of Mn$^{2+}$ ($\lambda_{em}$ = 670 nm). PLE spectrum for $\lambda_{em}$ = 770 nm has the same features as PLE spectrum for $\lambda_{em}$ = 670, but with much lower intensity (see Fig. S5 for detail). Nevertheless, even direct Mn$^{2+}$ excitation at 445 nm ($^6A_{1g}$ ($^6S$) → $^4T_{2g}$ ($^4D$)) in Li$_2$MnCl$_4$:Eu results in much broader emission band compared to undoped Li$_2$MnCl$_4$.

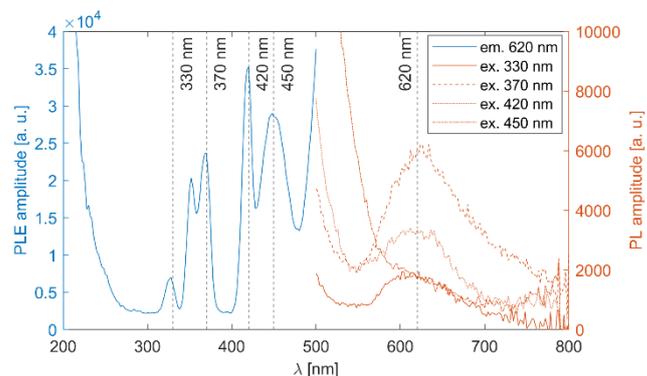

Figure 6: PLE and PL spectra of the undoped Li$_2$MnCl$_4$ sample. The excitation and emission wavelengths are indicated by dashed lines. The high intensity around 200 and 500 nm is due to overcorrection and light scattering respectively.

### Li$_2$MnCl$_4$:Ce$^{3+}$

Emission band which could be ascribed to 5d→4f emission of Ce$^{3+}$ was not observed in the PL spectra of Li$_2$MnCl$_4$:Ce. This is probably due to efficient energy transfer from Ce$^{3+}$ to Mn$^{2+}$. Based on the linear relationship between position of 5d→4f emission of Eu$^{2+}$ and Ce$^{3+}$ (see Dorenbos [20]) and the position of the lowest 4f-5d$_1$ absorption bands of $Eu^{2+}$ and $Ce^{3+}$ in chloride hosts[19,21] the 5d→4f emission of $Ce^{3+}$ in $Li_2MnCl_4$ should be centred around 320-340 nm which would result in strong overlap with the $^6A_{1g}$ ($^6S$) → $^4T_{1g}$ ($^4P$) absorption band of $Mn^{2+}$ enabling thus an efficient $Ce^{3+}$-$Mn^{2+}$ energy transfer further intensified by about 10x higher oscillator strength of 4f-5d1 transition of Ce3+ compared to that of $Eu^{2+}$. Similarly to the $Li_2MnCl_4$:Eu, the $Ce^{3+}$ doping gives rise to a broad emission band (see Fig. 8) extending from red to NIR. However, in contrast to $Eu^{2+}$ doping, the PL band in $Ce^{3+}$ doped sample does not exhibit maximum in the measured range. The $Ce^{3+}$ doping also results in appearance of excitation bands <320 nm which are most probably due to 4f-5d absorption bands of $Ce^{3+}$. However, in contrast to $Eu^{2+}$ this effect is significant only for emission in the deep red part of the emission band ($\lambda_{em}$ = 770).

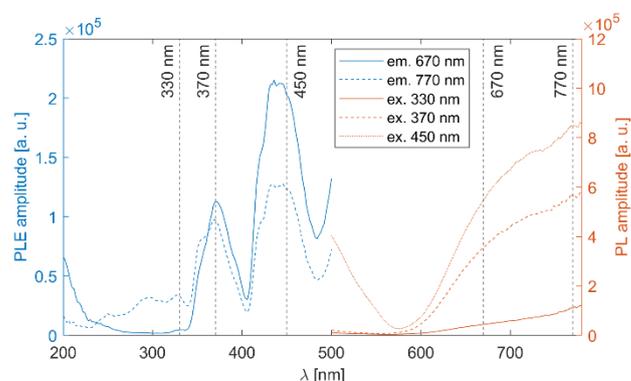

Figure 8: PLE and PL spectra of the Li$_2$MnCl$_4$:Ce$^{3+}$ (1 %) sample. The excitation and emission wavelengths are indicated by dashed lines. The high intensity around 200 and 500 nm is due to overcorrection and light scattering respectively.

## Photoluminescence decay kinetics

Photoluminescence decays were evaluated to get a better insight into luminescence mechanism in $Li_2MnCl_4$:X (X = $Eu^{2+}$, $Ce^{3+}$).

### $Li_2MnCl_4$

The PL decays of the broad band centred around 620 nm in undoped $Li_2MnCl_4$ were measured for different excitation wavelengths ($\lambda_{ex}$ = 330, 370, 420, and 450 nm). In all cases three exponential components were necessary for a satisfactory fit in the whole window (See Fig. S6). However, the first component with a decay time below resolution of the measurement (1.334 μs/ch) originates from light scattering and is included only to achieve satisfactory fit. Therefore, only two components correspond to $Mn^{2+}$ luminescence. Fast (~ 70 μs) and slow (~ 300 μs) components are present for all four excitation wavelengths with over 90 % contribution of the fast component (see Tab. S3). The intensity weighted mean decay times are 88.6, 87.5, 84.5, and 92.9 μs for excitation wavelengths 330, 370, 420, and 450 nm respectively. Similar values of the mean decay time agree with the assumption that the emission band in the undoped $Li_2MnCl_4$ results from $^4T_{1g}$ ($^4G$) → $^6A_{1g}$ ($^6S$) for all four excitation wavelengths. However, decay time around 1 ms is expected for the $^4T_{1g}$ ($^4G$) → $^6A_{1g}$ ($^6S$) transition due to violation of both parity and spin selection rules. The acceleration of the PL decay kinetics is most probably due to heavy concentration quenching which is possible when the luminescence centre is a matrix element. Such strong concentration quenching was reported by Song et al. [22] for heavily doped $Mn^{2+}$ doped spinel. Furthermore, a strong quenching would explain the fairly low intensity of $Mn^{2+}$ emission.

### $Li_2MnCl_4$:$Eu^{2+}$ and $Li_2MnCl_4$:$Ce^{3+}$

For $Li_2MnCl_4$:Eu the PL decay kinetics were monitored at three different emission wavelengths. At 405 nm corresponding to $Eu^{2+}$ 5d→4f emission and at 670 nm and 770 nm corresponding to the maximum and long wavelength edge of the red emission band respectively. The PL decay kinetics of the emission band centred at 405 nm, which was ascribed to 5d→4f emission of $Eu^{2+}$, was fitted using three exponential components (see Fig. 9).

The mean decay time is 53 ns. Such value is much lower than expected (and reported for $Eu^{2+}$ in chloride matrices [23–25]) for 5d→4f emission of $Eu^{2+}$ (~ 1 μs ) due to partial violation of the spin selection rule. However, the acceleration of PL decay kinetics is consistent with presence of energy transfer from $Eu^{2+}$ to $Mn^{2+}$. The PL decay kinetics of the red emission band was monitored at 670 nm and 770 nm for both the $Eu^{2+}$ and $Ce^{3+}$ doped samples. The excitation wavelengths were 340, 370, and 450 nm for $Li_2MnCl_4$:Eu (see Fig. S7) and 370 and 450 for $Li_2MnCl_4$:Ce (see Fig. S8). The results are summarized in Tab. 2. The PL decay times are longer for Ce doping. For both Eu and Ce doping the PL decay times are longer for the emission at 770 nm compared to 670 nm. The results suggest that the red emission band in doped $Li_2MnCl_4$ is composed of at least two components with different decay kinetics. The prolongation of the decay times is attributed to (i) increase of emission lifetime with emission wavelength[26] and (ii) partial suppression of concentration quenching resulting from larger Stokes shift.

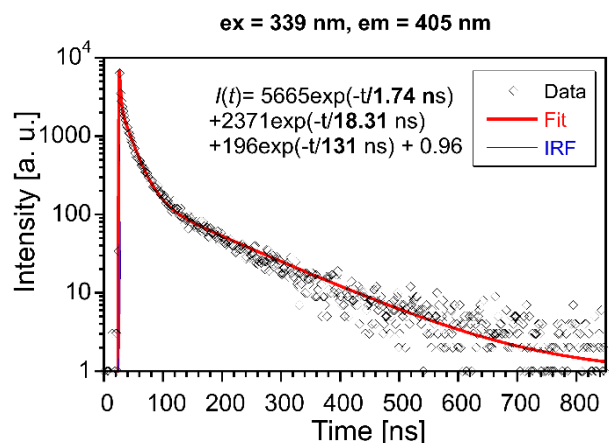

Figure 9: PL decay curve of $Li_2MnCl_4$:Eu ($\lambda_{ex}$ = 339 nm, $\lambda_{em}$ = 405 nm). Red solid line is convolution of function $I(t)$ and instrumental response function IRF.

Table 2: Mean PL decay times of Eu and Ce doped $Li_2MnCl_4$.

|  | Mean decay time [µs] | | | | |
| --- | --- | --- | --- | --- | --- |
|  | $Li_2MnCl_4$:Eu | | | $Li_2MnCl_4$:Ce | |
| em\ex | 340 nm | 370 nm | 450 nm | 370 nm | 450 nm |
| 670 nm | 129 | 131 | 115 | 172 | 171 |
| 770 nm | 171 | 162 | 151 | 209 | 215 |

**Radioluminescence spectra**

Similarly to the PL measurements, the RL emission of the doped samples exceeds the sensitive region of used PMT. Therefore, the RL spectra were measured using both PMT and CCD -based photodetectors (see Fig. S9). Figure 10 depicts the RL spectra of all three samples in the range from 200 to 1100 nm together with $Bi_4Ge_3O_{12}$ reference sample. The spectra were constructed by "gluing" data from PMT (200 – 800 nm) and CCD (400 – 1100 nm). The RL peak amplitude of undoped $Li_2MnCl_4$ was used to calculate amplitude ratios of both RL spectra (measured with PMT and CCD), because it is well reproduced in both ranges. The RL spectrum of $Li_2MnCl_4$:Eu shows a low intensity band centred at 405 nm (see Fig 10 and S10). The position coincides very well with PL band of $Eu^{2+}$ 5d→4f emission. A very weak structured band centred around 350 nm is observed in all three samples (see Fig S10). In Ce doped sample it seems to be overlapped with a different broad band. This could be due to 5d→4f emission of $Ce^{3+}$. However, the extremely low intensity does not allow precise assignment of either of the bands. The RL efficiencies (integral of RL spectra) are 1.2, 6.6, and 12.2 % of BGO reference sample for undoped $Li_2MnCl_4$, $Li_2MnCl_4$:Eu, and $Li_2MnCl_4$:Ce respectively.

The RL spectra of all three samples were fitted assuming gaussian band shape (see Fig. S11-13). For undoped $Li_2MnCl_4$ the spectrum is well fitted using single gaussian centred at 1.97 eV (629 nm). This is in agreement with assumption of single emission pathway $^4T_{1g}$ $(^4G)$ → $^6A_{1g}$ $(^6S)$. On the other hand, two gaussians were necessary for satisfactory fit of RL spectra of Eu and Ce doped samples (see Fig. S12 and S13). The centres of the gaussians coincide very well for both Eu (1.88 eV and 1.76 eV) and Ce (1.88 eV and 1.74 eV) doped samples. However, their relative contribution differs significantly. For Eu doped sample the contributions are 32.4 % and 67.6 % for the 1.88 eV (660 nm) and 1.76 eV (705 nm) respectively. While for the Ce doped sample the contributions are 12.4 % and 87.6 % for the 1.88 eV (660 nm) and 1.74 eV (713 nm) respectively. This would suggest that the nature of the luminescence centre is similar in both Eu and Ce doped samples. However, in Ce doped sample the lower energy centre is excited more efficiently.

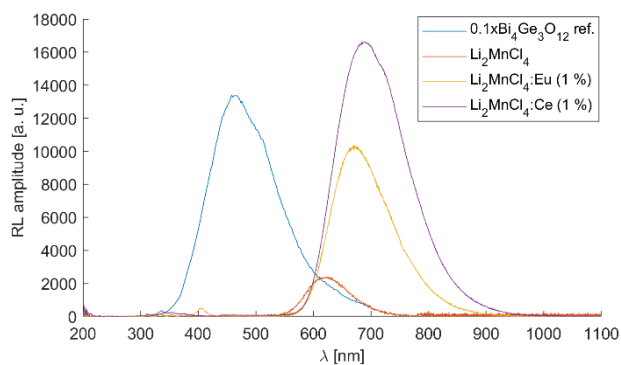

Figure 10: RL spectra of all three $Li_2MnCl_4$ samples together with BGO reference sample. The amplitude of BGO spectrum was divided by 10 for better comparison.

Based on the presented results we suggest a hypothesis explaining luminescence mechanisms in the $Eu^{2+}$ and $Ce^{3+}$ doped $Li_2MnCl_4$. The structure of $Li_2MnCl_4$ has two cationic crystallographic positions: tetrahedral occupied by Li and octahedral equally occupied by $Li^+$ and $Mn^{2+}$ (see XRPD results above). Due to larger size, the $Eu^{2+}$ and $Ce^{3+}$ cations should preferentially occupy the octahedral position. The crystal radius of $Eu^{2+}$ (1.31 Å) and $Ce^{3+}$ (1.15 Å) is 35 and 18 % larger than $Mn^{2+}$ (0.97 Å) in 6 coordination respectively. Therefore, the substitution of $Eu^{2+}$ or $Ce^{3+}$ for $Mn^{2+}$ (possible even octahedrally coordinated $Li^+$) results in expansion of the substituted octahedra and consequentially compression of its neighbouring $MnCl_6$ octahedra. Such compression results in red shift of the $^4T_{1g}$ ($^4G$) → $^6A_{1g}$ ($^6S$) emission due to stronger crystal field experienced by $Mn^{2+}$ (see Fig. 5). The emission from compressed $MnCl_6$ octahedra is less susceptible to concentration quenching due to larger Stokes shift. This effect could be further increased by substitution of $Cl^-$ with $S^{2-}$ from the starting materials (EuS or $Ce_2S_3$). Higher negative charge of sulphur compared to chlorine increases the crystal field and therefore contributes to the red shift of the $Mn^{2+}$ emission. Even though the segregation coefficient of sulphur should be very low it could incorporate into the lattice as a charge compensation centre i. e. following charge compensated octahedra could be present in the $Eu^{2+}$ or $Ce^{3+}$ $Li_2MnCl_4$: $(Eu_{Li}Cl_5S)^{5-}$, $(Eu_{Mn}Cl_6)^{4-}$, $(Ce_{Li}Cl_4S_2)^{5-}$, $(Ce_{Mn}Cl_5S)^{4-}$. If such charge compensation effect occurs a higher concentration of $S^{2-}$ is expected for $Ce^{3+}$ doped samples. Of course, other charge compensating defects like lithium vacancy or $Cl^-$ interstitial could also occur in the doped samples. The proposed effect together with efficient energy transfer from $Eu^{2+}$ and $Ce^{3+}$ to $Mn^{2+}$ would explain both redshift and increased intensity of $Mn^{2+}$ emission in $Eu^{2+}$ and $Ce^{3+}$ doped $Li_2MnCl_4$. However, further study is necessary to confirm or disprove incorporation of sulphur into $Li_2MnCl_4$.

Even though the PL decay kinetics of $Mn^{2+}$ in $Li_2MnCl_4$ is too slow to be used as a scintillator in a photon counting applications its emission matches very well an interval 650 – 1000 nm which was identified by Matsukura et al. [5] as target interval for emission of scintillation for long distance dose monitoring. This technique is being developed to monitor dose in harsh conditions, e. g. during decommissioning of Fukushima nuclear powerplant site. The $Li_2MnCl_4$:Ce could be used in a similar system for real time monitoring of neutron flux. Encapsulation of the $Li_2MnCl_4$:Ce in a polymer with high hydrogen content could serve three purposes at once: a protective casing to prevent air exposure, a neutron moderator, and a light guide to limit light losses. Moreover, low density ($\rho$ = 2.4 g/cm$^3$) and effective atomic number ($Z_{eff}$ = 17.1) of $Li_2MnCl_4$ will result in a very low gamma detection efficiency i. e. low gamma background. However, the feasibility of such a device would have to be tested in the high intensity mixed neutron gamma field ideally using fully enriched $^6Li$ to maximize neutron detection efficiency.

**Conclusions**

We presented a novel red emitting scintillator $Li_2MnCl_4$ which combines high Li content (28.5 %$_{at}$), low density ($\rho$ = 2.4 g/cm$^3$), and low effective atomic number ($Z_{eff}$ = 17.1). This makes $Li_2MnCl_4$ a promising candidate for remote neutron flux monitoring with low gamma background. Especially for measurements in high flux mixed neutron-gamma fields. The luminescence mechanism in undoped $Li_2MnCl_4$ was investigated in detail and all the absorption and emission bands were assigned according to Tanabe-Sugano diagram. Doping with $Eu^{2+}$ and $Ce^{3+}$ was investigated to improve the scintillation performance. The doping with $Eu^{2+}$ and $Ce^{3+}$ resulted in 5.5 and 10 times increase in radioluminescence efficiency respectively which points to the dopant-enhanced energy transfer towards Mn sublattice in scintillation mechanism taking into account that the increase of the photoluminescence mean decay time of $Mn^{2+}$ is less than 2.5 in doped samples compared to the undoped one (Table 2 and Table S3). However, even for the best sample the RL efficiency is only 12.2 % of $Bi_4Ge_3O_{12}$ reference sample which is presumably due to strong concentration quenching in the Mn-sublattice. The $Mn^{2+}$ emission lifetime of about 21-22 ms[17] was measured in $CaCl_2$ host at room temperature which is about 107x longer than the mean decay time measured for the Ce-doped $Li_2MnCl_4$ sample (Table 2). In the absence of concentration quenching, we could thus expect the integral efficiency of 107 x 12.2% > 1300% of BGO. Considering typical BGO scintillation light yield of 8000ph/MeV it provides an intrinsic light yield limit of $Li_2MnCl_4$-based material above 100 000 ph/MeV.

Based on the presented data we propose a luminescence mechanism in $Eu^{2+}$ and $Ce^{3+}$ doped

Li$_2$MnCl$_4$ involving energy transfer from the dopants to Mn$^{2+}$, a local lattice distortion around the dopant and a possible charge compensation in case of Ce$^{3+}$ dopant. Distorted/perturbed Mn$^{2+}$ sites nearby dopants show more red-shifted emission and lower susceptibility to concentration quenching. It consistently results in higher Mn$^{2+}$ emission intensity in RL spectra and slower PL decay. Concentration quenching in the Mn$^{2+}$ sublattice appears the main reason of rather low scintillation efficiency of Li$_2$MnCl$_4$-based materials in this study and future work will be focused onto diminishing this deteriorating process by suitable near infrared emitting dopants.

## Author Contributions

Vojtěch Vaněček – Conceptualization, Data curation, Formal Analysis, Investigation, Validation, Visualization, Writing – original draft, Writing – review & editing

Robert Král - Conceptualization, Funding acquisition, Project administration, Investigation, Validation, Supervision, Writing – review & editing

Kateřina Křehlíková - Data curation, Formal Analysis, Investigation, Validation, Visualization, Writing – review & editing

Kučerková Romana - Data curation, Formal Analysis, Investigation, Visualization

Vladimir Babin - Data curation, Formal Analysis, Investigation, Visualization

Petra Zemenová - Data curation, Formal Analysis, Investigation

Jan Rohlíček - Data curation, Formal Analysis, Investigation

Málková Zuzana - Data curation, Formal Analysis, Investigation

Jurkovičová Terézia - Data curation, Formal Analysis, Investigation

Martin Nikl - Conceptualization, Funding acquisition, Project administration, Supervision, Writing – review & editing

## Conflicts of interest

There are no conflicts to declare.

## Acknowledgements

The work is supported by Operational Programme Johannes Amos Comenius financed by European Structural and Investment Funds and the Czech Ministry of Education, Youth and Sports (Project No. SENDISO - CZ.02.01.01/00/22_008/0004596. This work was supported by JSPS KAKENHI Grant Number K24KF00040. This research was conducted in the scope of the Japanese Society for Promotion of Science standard fellowship.

## Notes and references